\documentclass[12pt]{article}
\usepackage{epsfig,rotating}
\begin{document}

\begin{titlepage}
\pagenumbering{arabic}
\vspace*{1.5cm}

\begin{center}

{\Large \bf Experimental study of direct photon emission\\
in \boldmath{$K^- \to \pi^-\, \pi^0\, \gamma$} decay using ISTRA+ detector}

\vspace*{1.0cm}
\normalsize
{V.A.~Uvarov, S.A.~Akimenko, G.I.~Britvich, K.V.~Datsko, A.P.~Filin,
A.V.~Inyakin, V.A.~Khmelnikov, A.S.~Konstantinov, V.F.~Konstantinov,
I.Y.~Korolkov, V.M.~Leontiev, V.P.~Novikov, V.F.~Obraztsov, V.A.~Polyakov,
V.I.~Romanovsky, V.M.~Ronjin, V.I.~Shelikhov, N.E.~Smirnov, O.G.~Tchikilev,
O.P.~Yushchenko}
  
\vskip 0.15cm
{\it Institute for High Energy Physics, Protvino, Russia}

\vskip 0.5cm
{V.N.~Bolotov, V.A.~Duk, S.V.~Laptev, A.Yu.~Polyarush}

\vskip 0.15cm
{\it Institute for Nuclear Research, Moscow, Russia}

\end{center}

\vspace*{2.0cm}

\noindent
{\small 
{\bf Abstract}

\vskip 0.15cm
The branching ratio in the charged-pion kinetic energy region of 55 to 
90 MeV for the direct photon emission in the $K^- \to \pi^- \pi^0 \gamma$ decay 
has been measured using in-flight decays detected with the ISTRA+ setup 
operating in the 25~GeV/$c$ negative secondary beam of the U-70 PS.
The value 
$$Br(DE) ~=~ 
\left[0.37 \pm 0.39\,({\rm stat}) \pm 0.10\,({\rm syst})\right]\times 10^{-5}$$
obtained from the analysis of 930 completely reconstructed events is consistent
with the average value of two stopped-kaon experiments, but it differs by 2.5 
standard deviations from the average value of three in-flight-kaon experiments.
The result is also compared with recent theoretical predictions.}

\end{titlepage}

\newpage
\section{Introduction}

The radiative decay channel $K^- \to \pi^- \pi^0 \gamma$\,
($K_{\pi 2 \gamma}$) is one of the most sensitive and important channels 
in investigating the chiral anomaly \cite{cha1} in the non-leptonic sector 
\cite{cha2,cha3}. This chiral anomaly is a basic feature of quantum field 
theories with chiral fermions and thus of the Standard Model.
Therefore, experimental tests of the chiral anomaly are crucial for the
theoretical basis of particle physics.

The total amplitude for the $K_{\pi 2 \gamma}$ decay can be generally 
decomposed as the sum of two terms: the inner bremsstrahlung (IB) 
associated with the decay $K^- \to \pi^- \pi^0$\, ($K_{\pi 2}$) 
in which the photon is emitted from the outgoing charged pion, and the direct 
emission (DE) in which the photon is emitted from one of the intermediate
states of the decay.
The inner bremsstrahlung is completely predicted by quantum electrodynamics 
in terms of the $K_{\pi 2}$ amplitude \cite{Low}. 
Because the $K_{\pi 2}$ decay is suppressed by the $\Delta I = 1/2$ 
isospin selection rule,
the bremsstrahlung contribution to the $K_{\pi 2 \gamma}$ decay is 
also suppressed. Unlike the bremsstrahlung,
the direct emission processes are permitted here, since for them
weak-electromagnetic transitions are possible from the initial kaon
to the final two-pion $P$-wave state with $I = 1$.
Although the inner bremsstrahlung component is still dominant, it can be 
isolated kinematically, 
and therefore the direct emission contribution can compete with the suppressed 
inner bremsstrahlung one.

The simplest radiative transitions for the direct emission in the 
$K_{\pi 2 \gamma}$ decay are electric $E1$ and magnetic 
$M1$ dipole transitions \cite{Good}. 
The electric transition can interfere with the inner bremsstrahlung process,
and possible non-standard-model effects, like a $CP$-violating asymmetry 
between $K^+_{\pi 2 \gamma}$ and $K^-_{\pi 2 \gamma}$ decay rates, 
could appear in the corresponding interference term. 
The magnetic transition is a manifestation of the chiral anomaly. 

Within the framework of Chiral Perturbation Theory (ChPT) at leading order,
$O(p^4)$, the magnetic transition amplitude consists of two different classes 
of anomalous amplitudes: reducible \cite{cha2} and direct \cite{cha3,cha4} 
amplitudes. The reducible anomalous amplitude is derived directly from the
Wess--Zumino--Witten functional \cite{WZW} and does not depend on undetermined 
constants, whereas the direct anomalous amplitude, arising from 
higher-dimension operators in ChPT, and also the electric transition amplitude 
are subject to some theoretical uncertainties.

The differential rate for the $K_{\pi 2 \gamma}$ decay is 
conveniently expressed in terms of the Dalitz plot variables $T^*_c$ and $W$, 
where $T^*_c$ is the kinetic energy of $\pi^-$ in the $K^-$ rest frame,
and $W^2 \equiv (p\cdot q)\,(p_c\cdot q)\,/\,(m^2_{\pi^-}\,m^2_{K^-})$.\,
Here $p$,\, $p_c$ and $q$ are 4-momenta of $K^-$, $\pi^-$ and
$\gamma$, and $m_{\pi^-}$ and $m_{K^-}$ are masses of $\pi^-$ and $K^-$,
respectively. This rate can be written \cite{HANDBOOK} in terms of the inner 
bremsstrahlung differential rate as
\begin{equation}
\label{eq1}
\frac{\partial^{\,2}\,\Gamma}{\partial\,T^*_c~\partial\,W}~=~
\frac{\partial^{\,2}\,\Gamma_{IB}}{\partial\,T^*_c~\partial\,W}\left\{
1+2\,\frac{m^2_{\pi^-}}{m_{K^-}}\,{\rm Re}\left[\frac{E}{eA}\right]W^2
+\,\frac{m^4_{\pi^-}}{m^2_{K^-}}\left( \left|\frac{E}{eA}\right|^2+
\left|\frac{M}{eA}\right|^2 \right)W^4 \right\},
\end{equation}
where $A$ is the on-shell amplitude for the $K_{\pi 2}$ decay,
and $E$ and $M$ are the direct emission electric and magnetic invariant
dimensionless amplitudes, defined in Ref. \cite{cha2}.

In experimental studies of the $K_{\pi 2 \gamma}$ decay the variable 
$T^*_c$ is usually used to minimize contaminations arising from the 
$K_{\pi 2}$ decay, dominated at $T^*_c > 90$ MeV, and from the decay 
$K^- \to \pi^- \pi^0 \pi^0$\, ($K_{\pi 3}$), dominated at 
$T^*_c < 55$ MeV.
The variable $W$ is convenient to isolate the inner bremsstrahlung from the 
direct emission, since the former process dominates at small values of $W$, 
while the latter dominates at large values of $W$. 

The branching ratio in the region of 55~MeV\,$< T^*_c <$~90~MeV for the 
direct photon emission in the $K^{\pm}_{\pi 2 \gamma}$ decay ($Br$) has been 
measured in the following five experiments: 
three in-flight-kaon and two stopped-kaon experiments have found the 
weighted average values of $Br$ = (1.8$\,\pm\,$0.4)$\times$10$^{-5}$ 
\cite{Abrams72,Smith76,Bolotov87} and
$Br$ = (0.44$\,\pm\,$0.08)$\times$10$^{-5}$ \cite{Adler00,Aliev03}, 
respectively. The discrepancy between these two average values
is 3.3 standard deviations. Therefore, further experimental studies 
are necessary to settle this discrepancy.

The cited experimental values of $Br$ can be compared with corresponding 
theoretical predictions in the same $T^*_c$ region.
Under the assumption that the direct emission is entirely
due to the reducible anomalous amplitude given by Eq.~(30b) in 
Ref. \cite{cha2} with standard $O(p^2)$ ChPT coupling constants,
ChPT at leading order predicts the value of $Br$ = 0.35$\times$10$^{-5}$. 
Model-dependent theoretical predictions which take into account the additional 
contribution of the direct anomalous amplitude \cite{cha3,cha4} are also 
available: for example, the factorization model \cite{cha4} predicts 
the value of $Br$ = 1.94$\times$10$^{-5}$.

The above-mentioned observations encourage us to perform a new measurement 
of the direct photon emission in the $K_{\pi 2 \gamma}$ decay 
using in-flight negative kaons.

\section{Experimental setup}

The experiment has been performed at the IHEP proton synchrotron U-70 with the
experimental apparatus ISTRA+, which is a modification of the ISTRA-M setup 
\cite{ISTRA-M} and which was described in some details in our recent papers 
where studies of the $K_{e3}^-$~\cite{plb1}, $K_{\mu 3}^-$~\cite{plb2} and 
$K_{\pi3}^-$~\cite{plb3} decays were presented. 
The setup is located in the negative unseparated secondary beam with the 
following parameters during the measurement: the momentum is $\sim25$~GeV/$c$ 
with $\sigma (p)/p \sim 1.5\,\%$, the admixture of kaons is $\sim3\,\%$, and 
the total intensity is $\sim3$$\times$10$^6$ per spill.

\begin{figure}
\begin{turn}{90}
\centering\mbox{\epsfig{file=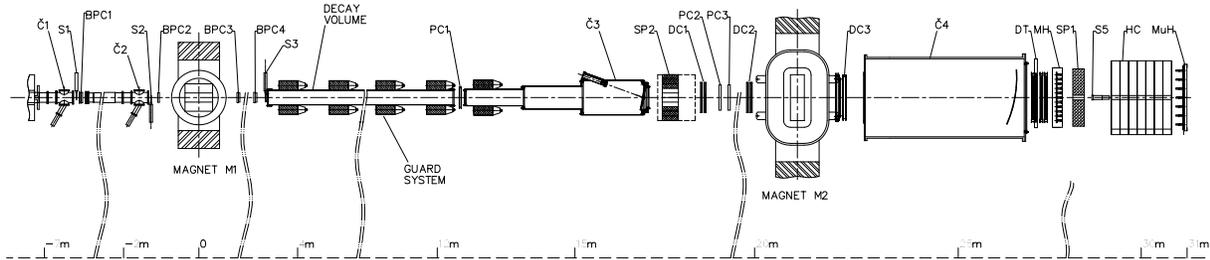,height=\textwidth}}
\end{turn}
\caption{\small The side elevation view of the ISTRA+ detector.}
\label{fig1}
\end{figure}

The side elevation view of the ISTRA+ detector is shown in Fig.\,\ref{fig1}. 
The setup coordinate system is the following: the $x$, $y$ and $z$ axes are 
turned along the field of the spectrometer magnet M2, the vertical line and 
the setup longitudinal axis, respectively.

The measurement of the beam particles, deflected by the beam magnet M1, 
is performed with four beam proportional chambers BPC1--BPC4. 
The kaon identification is done by three threshold gas Cherenkov counters 
$\check{\rm C}$0--$\check{\rm C}$2 
($\check{\rm C}$0 is not shown in Fig.\,\ref{fig1}). 
The momenta of the secondary charged particles, deflected in the vertical plane
by the spectrometer magnet M2, are measured with three proportional chambers 
PC1--PC3, three drift chambers DC1--DC3 and four planes of the drift tubes DT. 
The secondary photons are detected by the lead-glass electromagnetic 
calorimeters SP1 and SP2.
To veto low energy photons the decay volume is surrounded by the guard system 
of eight lead-glass rings and by the SP2. 
The wide aperture threshold helium Cherenkov counters $\check{\rm C}$3 and
$\check{\rm C}$4 are not used in the present study.
In Fig.\,\ref{fig1}, HC is a scintillator-iron sampling hadron calorimeter, 
MH is a scintillation hodoscope used to improve the time resolution of the 
tracking system, MuH is a scintillation muon hodoscope.

The trigger is provided by the scintillation counters S1--S5, the Cherenkov 
counters $\check{\rm C}$0--$\check{\rm C}$2 and the analog sum of the 
amplitudes from the last dinodes of the calorimeter SP1 
(see Refs.~\cite{plb1,plb2} for details). The latter serves to suppress 
the dominating $K^- \to \mu^- \bar{\nu}_{\mu}$ decay. 

\section{Event selection}

About 332\,M events were collected during one physics run in Winter 2001. 
These experimental data are complemented by about 
260\,M events generated with the Monte Carlo program GEANT3 \cite{geant}. 
The Monte Carlo simulation includes a realistic description of the 
experimental setup: the decay volume entrance windows, the track chamber 
windows, gas mixtures, sense wires and cathode structures, the Cherenkov 
counter mirrors and gas mixtures, the showers development in the 
electromagnetic calorimeters, etc. 
The details of the reconstruction procedure have been published in 
Refs.~\cite{plb1,plb2}, here only key points relevant to the 
$K^- \to \pi^-\pi^0\gamma$ event selection are described.

The data processing starts with the beam track reconstruction in the 
beam proportional chambers BPC1--BPC4, and then with the secondary tracks 
reconstruction in the decay tracking system PC1--PC3, DC1--DC3 and DT. 
The decay vertex is reconstructed by means of the unconstrained vertex fit
of the beam and decay tracks. Finally, the electromagnetic showers are looked 
for in the calorimeters SP1 and SP2, and the photons are reconstructed
using the fit procedure with the Monte Carlo generated two-dimensional patterns 
of showers. To suppress leptonic $K^-$ decays the particle identification 
is used. 
The electrons are identified using the ratio of the energy of the shower, 
detected in the calorimeter SP1 and associated with the track of the electron,
to the momentum of the electron \cite{plb1}.
The muons are identified using the information from the calorimeters SP1 
and HC \cite{plb2}.

In the present study, the main purpose of the event selection is to suppress 
significantly all components of the background contamination, even if some of 
them are negligible in studying the $K^-$ decay modes which branching ratios
are more than 1\%. The expediency of the selection criteria, mentioned below, 
is motivated by the Monte Carlo investigation.

At the first step of the event selection only the measurements of the beam and 
secondary charged particles are used. Those events are selected which satisfy
the following requirements:
\begin{itemize}
\item[--]
only one beam track and one negative secondary track are detected;
\item[--]
the first hit of the secondary track is either in the chamber PC1, or in 
the DC1, or in the PC2, while the last hit of this track is in the drift 
tubes DT;
\item[--]
the probability of the vertex fit, $CL(\chi^2)$, is more than $10^{-4}$;
\item[--]
the relative error of the secondary track momentum, $\sigma (p)/p$, is less 
than 0.1;
\item[--]
the decay vertex is before the calorimeter SP2 (6~m $<z<$ 17~m), and its 
transverse position is in the region of ($-$3~cm $<x<$ 3~cm,
$-$2~cm $<y<$ 6~cm); 
\item[--]
the secondary track is not identified as an electron or as a muon;
\item[--]
the angle between the $K^-$ line of flight and the $\pi^-$ direction in the 
$K^-$ rest frame is in the region of $-0.8<\cos\theta^*_{\pi^-}<0.85$;
\item[--]
the $\pi^-$ kinetic energy in the $K^-$ rest frame is in the region 
of 55 to 90 MeV.
\end{itemize}

At the second step of the event selection the measurements of the showers in 
the calorimeters SP1 and SP2 are used. 
Associating the SP1 shower with the secondary track is done if the distance 
$R=[(x_{sh}-x_{tr})^2+(y_{sh}-y_{tr})^2]^{1/2}$ is less than 3 cm, where 
($x_{sh}$,\,$y_{sh}$) and ($x_{tr}$,\,$y_{tr}$) are the transverse coordinates 
of the shower and of the track extrapolation to the calorimeter SP1, 
respectively. The event selection is done by the requirements:
\begin{itemize}
\item[--]
the number of showers associated with the secondary track ($N_{ass}$) is no 
more than one, and if the event has such shower (i.e. $N_{ass}=1$) the ratio 
of the energy of this shower to the track momentum is less than 0.7;
\item[--]
the total (in both calorimeters) number of photons (i.e. showers, which are not 
associated with the secondary track) is equal to three.
\end{itemize}
Also, the events are selected if the three photons satisfy the following 
requirements:
\begin{itemize}
\item[--]
at least one of them is detected in the SP1; 
\item[--]
the photon energy is more than 0.8 GeV, but is more than 1 GeV if the photon 
is detected in the SP2;
\item[--]
the relative error of the photon energy, $\sigma (E)/E$, is less than 1, but
the number of photons with $\sigma (E)/E > 0.2$ is no more than one;
\item[--]
the relative transverse position of the SP1 photon
($\Delta x = x_{sh}-x_{tr}$, $\Delta y = y_{sh}-y_{tr}$) 
is outside the rectangle region of 
($|\Delta x| < 7$ cm, $|\Delta y| < 11$ cm) and, if $N_{ass}=1$,
the strip region of ($|\Delta x| < 2$ cm, $\Delta y \geq 11$ cm);
\item[--]
the photon configuration is not such as the one, in which all SP1 photons 
are reconstructed from one and the same cluster of overlapped showers.
\end{itemize}

At the third step of the event selection the hits found in the hodoscope MH,
in the proportional chamber PC1 and in the drift tubes DT are considered to 
minimize the background contamination and systematics. The further selection 
is done by the requirements: 
\begin{itemize}
\item[--]
the number of hits in the hodoscope MH ($n_{MH}$) is no less than one, 
and exactly one of them is associated with the secondary track extrapolation
to the hodoscope, all the others (if $n_{MH}>1$) are associated with the SP1
photon interpolations to the hodoscope (in terms of the distance in the MH 
transverse plane, $r < 10$ cm);
\item[--]
the number of hits, which are found in the chamber PC1, but not used in the 
track reconstruction, is no more than one for each coordinate plane;
\item[--]
if some hit found in the $y$-coordinate plane of the chamber PC1 is used in 
the secondary track reconstruction, the decay vertex is in the region before 
this chamber; 
\item[--]
any hit found in the DT $x$- or $y$-coordinate plane is either used in the 
track reconstruction or associated with at least one SP1 photon interpolation 
to the drift tubes (in terms of the distance in the DT coordinate plane, 
$\delta x < 1$~cm or $\delta y < 1$~cm). 
\end{itemize}

\begin{figure}
\centering\mbox{\epsfig{file=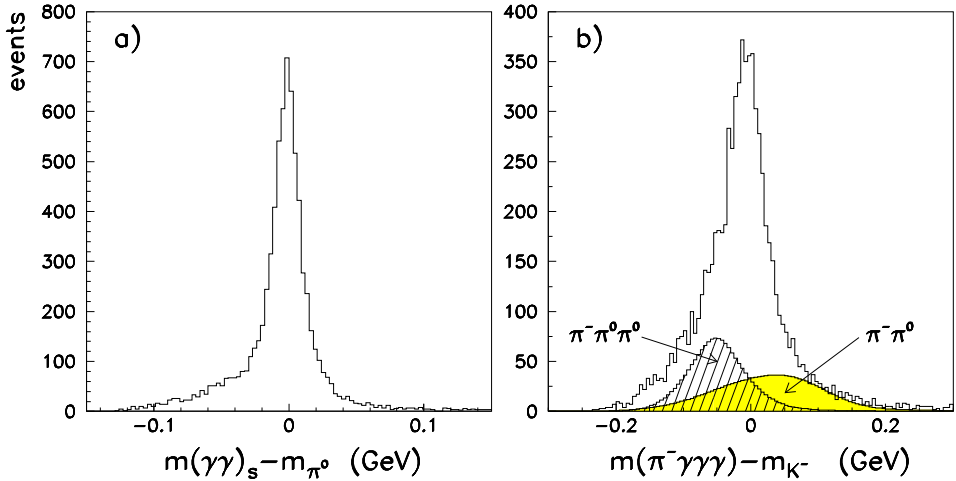,width=0.80\textwidth}}
\caption{\small 
Mass deviations in the events after the third step of the event selection:\,
{\bf a)}~the deviation from the $\pi^0$ mass, 
$m(\gamma\gamma)_s-m_{\pi^0}$, of the effective mass of the 
$\gamma\gamma$ pair with the smallest absolute value of the deviation;~
{\bf b)}~the deviation from the $K^-$ mass, 
$m(\pi^-\gamma\gamma\gamma)-m_{K^-}$, of the effective mass of the 
$\pi^-\gamma\gamma\gamma$ system, together with the Monte Carlo estimated 
background contaminations of the $K^- \to \pi^- \pi^0 \pi^0$ and 
$K^- \to \pi^- \pi^0$ events (hatched histograms).}
\label{fig2}
\end{figure}

To illustrate the quality of the first three steps of the event selection the
mass deviations $m(\gamma\gamma)_s-m_{\pi^0}$ and 
$m(\pi^-\gamma\gamma\gamma)-m_{K^-}$ for the corresponding events are shown in 
Fig.\,\ref{fig2}, where $m(\gamma\gamma)_s$ is the effective mass of the 
$\gamma\gamma$ pair with the smallest absolute value of the deviation,
$m(\pi^-\gamma\gamma\gamma)$ is the effective mass of the 
$\pi^-\gamma\gamma\gamma$ system, and $m_{\pi^0}$ is the $\pi^0$ mass. 
The background contamination of the $K^- \to \pi^- \pi^0 \pi^0$ and 
$K^- \to \pi^- \pi^0$ events (also shown in Fig.\,\ref{fig2}b) is estimated 
from the Monte Carlo simulation to be 41\%. 

At the fourth step of the event selection the kinematic criteria are 
used to select the $K^- \to \pi^- \pi^0 \gamma$ events. First, for each of 
the three $\gamma\gamma$ combinations in the event, assuming it arises from the
$\pi^0 \to \gamma\gamma$ decay, the kinematic 5C-fit for the 
$K^- \to \pi^- \pi^0 \gamma$ hypothesis is applied. 
Then, such $\pi^0 \to \gamma\gamma$ pairing is chosen, for which the event
passes the kinematic fit with the largest value of the combined probability
$P_c = P_{fit\,}P_{IB}$, where $P_{fit}$ is the $\chi^2$ probability of 
the fit and $P_{IB}$ is the inner bremsstrahlung decay probability as a 
function of the {\it fitted} values of the Dalitz plot variables.

\begin{figure}
\centering\mbox{\epsfig{file=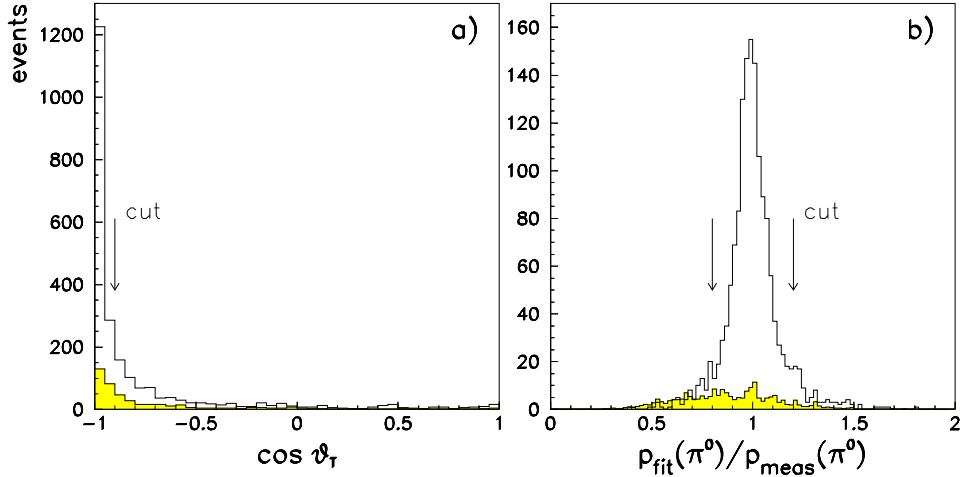,width=0.80\textwidth}}
\caption{\small 
Distributions of the events passed the kinematic 5C-fit for the 
$K^- \to \pi^- \pi^0 \gamma$ hypothesis, together with the Monte Carlo 
estimated background contaminations of the $K^- \to \pi^- \pi^0 \pi^0$ and
$K^- \to \pi^- \pi^0$ events (hatched histograms):\,
{\bf a)}~the angle, $\theta_T$, between the measured $\pi^-\pi^0$ and $\gamma$ 
transverse momenta defined with respect to the $K^-$ direction;~
{\bf b)}~the ratio, $p_{fit}(\pi^0)/p_{meas}(\pi^0)$, of the fitted $\pi^0$ 
momentum value to the measured one.}
\label{fig3}
\end{figure}

After that, the event selection is done restricting the allowed ranges of the
variables defined by the {\it measured} values of the particle momenta. 
The corresponding criteria, in which the measured $\pi^0$ four-momentum is 
multiplied (re-scaled) by a factor $\lambda=m_{\pi^0}/m(\gamma\gamma)$, are 
the following:
\begin{itemize}
\item[--]
the measured momentum of the $\pi^-\pi^0\gamma$ system is in the region of
23 to 29 GeV/$c$;
\item[--]
the angle between the measured $\pi^-\pi^0$ and $\gamma$ transverse momenta 
defined with respect to the $K^-$ direction is more than 154$^{\circ}$
(see Fig.\,\ref{fig3}a);
\item[--]
the ratios, 
$p_{fit}(\pi^0)/p_{meas}(\pi^0)$ and $p_{fit}(\gamma)/p_{meas}(\gamma)$, of the
fitted $\pi^0$ and $\gamma$ momenta values to the measured ones are in the 
region of 0.8 to 1.2 (see Fig.\,\ref{fig3}b);
\item[--]
the deviations
$|m(\gamma\gamma)-m_{\pi^0}|$ and $|m(\pi^-\pi^0\gamma)-m_{K^-}|$
are less than 40 MeV/$c^2$;
\end{itemize}
where $m(\gamma\gamma)$ is the effective mass of the $\gamma\gamma$ pair arising
from the $\pi^0$ decay, $m(\pi^-\pi^0\gamma)$ is the effective mass of the 
$\pi^-\pi^0\gamma$ system. 

\begin{figure}
\centering\mbox{\epsfig{file=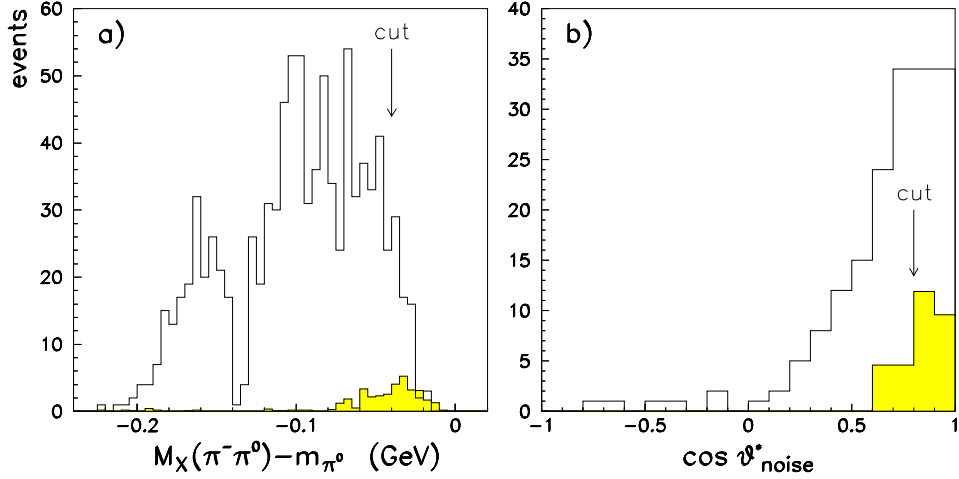,width=0.80\textwidth}}
\caption{\small 
Distributions of the events before the last cut of the fifth step of the event 
selection, together with the Monte Carlo estimated background contaminations
(hatched histograms):\,
{\bf a)}~for the $K^- \to \pi^- \pi^0 \pi^0$ hypothesis, the deviation from the
$\pi^0$ mass, $M_X(\pi^- \pi^0)- m_{\pi^0}$, of the missing mass to the 
$\pi^- \pi^0$ system with the reconstructed and re-scaled $\pi^0$ meson;~ 
{\bf b)}~for the $K^- \to \pi^- \pi^0$ hypothesis, the angle in the $K^-$ rest 
frame, $\theta^*_{noise}$, between the $\pi^-$ and $\gamma$ (considered as a 
calorimeter noise) momenta.}
\label{fig4}
\end{figure}

At the fifth step of the event selection the contaminations of the 
$K^- \to \pi^- \pi^0 \pi^0$ and $K^- \to \pi^- \pi^0$ events are minimized.
The event is not selected if, for at least one $\pi^0 \to \gamma\gamma$ 
pairing, it passes the kinematic fit for the $K^- \to \pi^- \pi^0 \pi^0$ 
(one $\gamma$ in the event is lost) or $K^- \to \pi^- \pi^0$ (one $\gamma$ in
the event is a calorimeter noise) hypothesis and satisfies the additional
requirements, which for the $K^- \to \pi^- \pi^0 \pi^0$ hypothesis are: 
\begin{itemize}
\item[--]
the Dalitz plot variable $W$ (defined for the chosen above 
$K^- \to \pi^- \pi^0 \gamma$ hypothesis) is more than 0.25;
\item[--]
the deviation $|m(\pi^- \pi^0 \pi^0)- m_{K^-}|$ is less than 40 MeV/$c^2$,
where $m(\pi^- \pi^0 \pi^0)$ is the effective mass of the $\pi^- \pi^0 \pi^0$
system, in which the first $\pi^0$ is reconstructed from the measured photons 
and re-scaled as above by a factor $\lambda$, while the second one is 
calculated from the balance of the three-momenta;
\item[--]
the deviation $|M_X(\pi^- \pi^0)- m_{\pi^0}|$ is less than 40 MeV/$c^2$, where 
$M_X(\pi^- \pi^0)$ is the missing mass to the $\pi^- \pi^0$ system with the 
reconstructed and re-scaled $\pi^0$ meson (see Fig.\,\ref{fig4}a); 
\end{itemize}
and which for the $K^- \to \pi^- \pi^0$ hypothesis are: 
\begin{itemize}
\item[--]
the angle between the measured $\pi^-$ and $\pi^0$ transverse momenta defined
with respect to the $K^-$ direction is more than 174$^{\circ}$;
\item[--]
the deviation $|m(\pi^- \pi^0)- m_{K^-}|$ is less than 50 MeV/$c^2$, where 
$m(\pi^- \pi^0)$ is the effective mass of the $\pi^- \pi^0$ system with 
the re-scaled $\pi^0$ meson; 
\item[--]
the angle in the $K^-$ rest frame between the $\pi^-$ and $\gamma$ (the latter 
is considered here as a calorimeter noise) momenta is less than 37$^{\circ}$
(see Fig.\,\ref{fig4}b).
\end{itemize}

Using the mentioned above selection criteria for the $K^- \to \pi^-\pi^0\gamma$
decay we have collected 930 completely reconstructed events. 
The corresponding numbers of accepted Monte Carlo events are about 13 times 
larger than the ones collected in the experiment. 
The surviving background contamination arising from all background
decay modes is estimated from the Monte Carlo simulation to be less than 6\%. 
The detailed event reduction statistics is given in Table~\ref{tab1}. 
\begin{table}[bth]
\caption{The event reduction statistics.}
\renewcommand{\arraystretch}{1.5}
\begin{center}
\begin{tabular}{|l|r|}
\hline
Total number of events& 332\,M \\
\hline
Beam track reconstructed& 248\,M \\
\hline
Secondary track(s) reconstructed& 124\,M \\                    
\hline
Number of events written on DST& 108\,M \\
\hline \hline
$K^-$ and $\pi^-$ selected (all cuts of the 1st step)& 1729\,K \\
\hline
$\gamma\gamma\gamma$ selected (all cuts of the 2nd and the 3rd steps)& 7623 \\
\hline
$K^- \to \pi^-\pi^0\gamma$ pre-selected (all cuts of the 4th step)& 1041 \\
\hline
$K^- \to \pi^-\pi^0\gamma$ selected (all cuts of the 5th step)& ~930 \\
\hline
\end{tabular}
\end{center}
\label{tab1}
\end{table}
\clearpage

\section{Analysis}

\begin{figure}
\centering\mbox{\epsfig{file=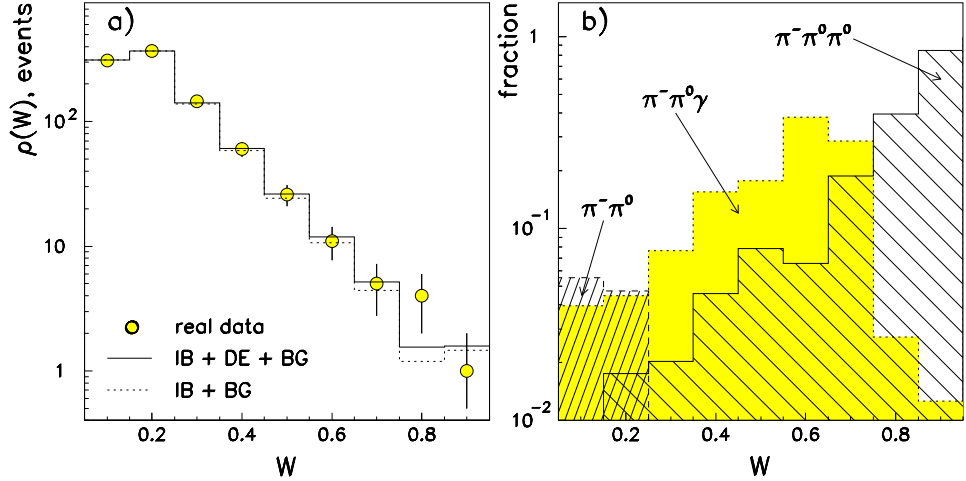,width=0.80\textwidth}}
\caption{\small 
Dependences on the Dalitz plot variable $W$:\,
{\bf a)}~the uncorrected number of the selected $K^- \to \pi^- \pi^0 \gamma$ 
events, together with the corresponding Monte Carlo estimations (the full line
is the sum of the IB, DE and BG components, but the dotted line is the sum of 
the IB and BG ones);~
{\bf b)}~the Monte Carlo estimated fractions in the selected 
$K^- \to \pi^- \pi^0 \gamma$ events of the contaminations 
arising from the background $K^- \to \pi^- \pi^0$ and 
$K^- \to \pi^- \pi^0 \pi^0$ decays and from the wrong reconstructed
$K^- \to \pi^- \pi^0 \gamma$ decay.}
\label{fig5}
\end{figure}

In the present analysis, the uncorrected distribution $\rho(W)$ of the 
selected $K^- \to \pi^- \pi^0 \gamma$ events as a function of the Dalitz plot 
variable $W$ was used. This distribution is shown in Fig.\,\ref{fig5}a together
with the normalized distribution of the Monte Carlo simulated events
reconstructed with the same program as for the real data. 
The background was estimated from the simulation of the particle interaction 
with the material of the detector and of the kaon decay including all decay 
modes with the branching ratios more than 1\%. The corresponding branching 
ratios and matrix elements in the Monte Carlo simulation were taken from the 
PDG \cite{PDG}. 

The Monte Carlo distribution in Fig.\,\ref{fig5}a includes the 
$K^- \to \pi^- \pi^0 \gamma$ component (with and without the direct emission) 
and the background contamination (BG).
The fractions in the selected $K^- \to \pi^- \pi^0 \gamma$ events 
of the contaminations arising from
the background $K^- \to \pi^- \pi^0$ and $K^- \to \pi^- \pi^0 \pi^0$ decays
are shown in Fig.\,\ref{fig5}b as a function of the variable $W$ and estimated
to be 4\% and 2\%, respectively. The background of all other decay modes is 
negligible. Due to the detector imperfection, the $K^- \to \pi^- \pi^0 \gamma$ 
component in the simulated events consists of the correct and wrong 
reconstructed $K^- \to \pi^- \pi^0 \gamma$ decays. The fraction of the latter 
is also shown in Fig.\,\ref{fig5}b and estimated to be 6\%.

\begin{figure}
\centering\mbox{\epsfig{file=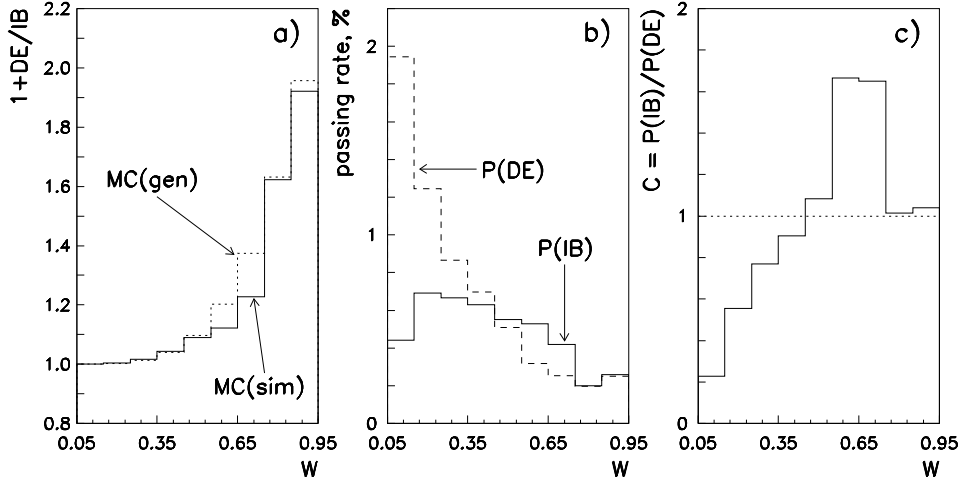,width=0.80\textwidth}}
\caption{\small 
Dependences on the variable $W$ estimated from the Monte Carlo
$K^- \to \pi^- \pi^0 \gamma$ events without the contamination
of the background decay modes:\,
{\bf a)}~the ratio 1+$DE/IB$ for the generated (dotted line) and for the 
simulated (full line) events;~
{\bf b)}~the event passing rates for the IB (full line) and for the DE 
(dashed line) components;~ 
{\bf c)}~the ratio of the passing rates for the IB and DE components.}
\label{fig6}
\end{figure}

Fig.\,\ref{fig6}a shows two $W$-dependences of the ratio of the DE and IB 
components estimated from the Monte Carlo $K^- \to \pi^- \pi^0 \gamma$ events 
without the contamination of the background decay modes. 
The first dependence was obtained from the {\it generated} events taken 
``before passing the detector'', while the second one was obtained from the 
{\it simulated} events taken ``after passing the detector''. 
Comparing these dependences, one can conclude that in the present experiment 
the DE component is not smeared due to the detector imperfection:
the acceptance and inefficiency, the event selection, the experimental 
resolution, the noise, the secondary interactions and the wrong 
$\pi^0 \to \gamma\gamma$ pairing.

The passing rate for the $K^- \to \pi^- \pi^0 \gamma$ events,
$P = N_{sim}/N_{gen}$, is shown in Fig.\,\ref{fig6}b as a function of the 
variable $W$ separately for the IB and DE components. These dependences were 
obtained from the Monte Carlo $K^- \to \pi^- \pi^0 \gamma$ events without the 
contamination of the background decay modes:\, $N_{gen}$ ($N_{sim}$) is the 
number of the generated (simulated) events, and the binning is given by the 
generated (simulated) value of the variable $W$. Fig.\,\ref{fig6}c shows the 
ratio of the passing rates for the IB and DE components, 
\begin{equation}
\label{eq2}
C = \frac{P(IB)}{P(DE)}. 
\end{equation}

To determine the DE amplitude $M$ in Eq.\,(\ref{eq1}) (under the assumption 
that the DE amplitude $E$ in Eq.\,(\ref{eq1}) is equal to zero) the 
experimental distribution $\rho(W)$ was fitted by the method of least squares 
with the function 
\begin{equation}
\label{eq3}
\rho(W)_{fit} ~=~ 
\alpha\left[\rho(W)_{\rm BG} + \rho(W)_{\rm IB} + \beta\rho(W)_{\rm DE}\right],
\end{equation}
where $\rho(W)_{\rm BG}$ is the background contamination obtained from the 
Monte Carlo simulated $K^- \to \pi^- \pi^0$ and $K^- \to \pi^- \pi^0 \pi^0$ 
events only,\, $\rho(W)_{\rm IB}$ and $\rho(W)_{\rm DE}$ are the IB and DE 
components obtained from the Monte Carlo simulated 
$K^- \to \pi^- \pi^0 \gamma$ events without the contamination of the background
decay modes, $\alpha$ and $\beta$ are free parameters in the fit.
This method allows to avoid the systematic errors \cite{anikeev} due to the 
``migration'' of the events on the Dalitz plot because of the finite 
experimental resolution.

\section{Results}

\begin{figure}
\centering\mbox{\epsfig{file=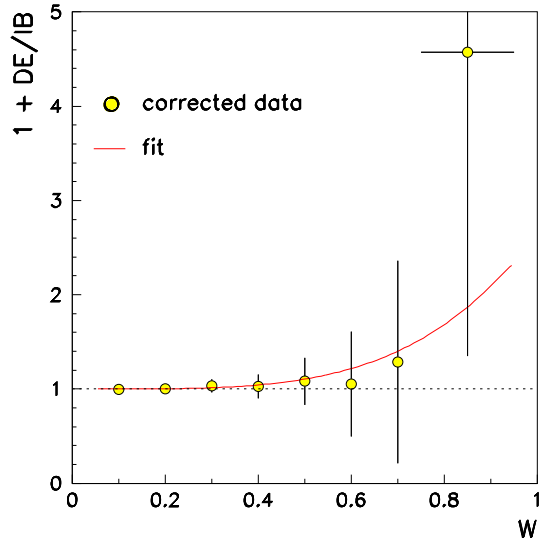,width=0.45\textwidth}}
\caption{\small 
The corrected experimental ratio of the DE and IB components for the 
$K^- \to \pi^-\pi^0\gamma$ decay as a function of $W$ in the region of
55~MeV\,$< T^*_c <$~90~MeV. The curve is the prediction of Eq.\,(\ref{eq1}) 
with the fitted value of the amplitude $M$ and with $E=0$.}
\label{fig7}
\end{figure}

The result of the least squares fit of Eq.\,(\ref{eq3}) to the distribution
$\rho(W)$ is illustrated in Fig.\,\ref{fig7}, where the corrected experimental 
ratio 
\begin{equation}
\label{eq4}
R(W)_{exp} ~=~ 1 +
C\cdot\left[\frac{\rho(W)-\alpha\,\rho(W)_{\rm BG}}{\alpha\,\rho(W)_{\rm IB}}
- 1\right]
\end{equation}
of the DE and IB components for the $K^- \to \pi^-\pi^0\gamma$ decay 
is shown as a function of $W$ in the $T^*_c$ region of 55 to 90 MeV.
In Eq.\,(\ref{eq4}), the correction factor $C$ given by Eq.\,(\ref{eq2})
takes into account the detector imperfection, the $\alpha\,\rho(W)_{\rm BG}$ 
term takes into account the contamination of the background decay modes,
and the fitted value of $\alpha$ provides the normalization.
The curve in Fig.\,\ref{fig7} is predicted by Eq.\,(\ref{eq1}) with
the fitted value of the amplitude $M$ (given by the fitted value of $\beta$) 
and with $E=0$.

From the least squares fit with the value of $\chi^2/ndf = 2.1/7$, the
direct emission component was obtained to be
\begin{center}
$|M| ~=~ (1.9 \pm 1.0 \pm 0.3)\times 10^{-7}$,
\end{center}
\begin{center}
$Br(DE)/Br(IB) ~=~ (1.4 \pm 1.5 \pm 0.4)$\,\%
\end{center}
under the assumption that there is no interference component. Then, comparing 
this with the theoretical value of the inner bremsstrahlung branching ratio, 
$Br(IB)$ = 2.61$\times$10$^{-4}$ \cite{HANDBOOK}, the direct emission 
branching ratio was determined to be
\begin{center}
$Br(DE) ~=~ (0.37 \pm 0.39 \pm 0.10)\times 10^{-5}$
\end{center}
for the $K^- \to \pi^-\pi^0\gamma$ decay in the region of
55~MeV\,$< T^*_c <$~90~MeV. Here the first errors are statistical and the 
second ones are systematic.

In the determination of the systematic uncertainty of $Br(DE)$ the following 
sources of systematics were investigated.
\begin{itemize}
\item[--]
The branching ratios and matrix elements of the background decay modes used in 
the Monte Carlo simulation were varied within their errors ($\Delta Br$ = 0.01).
\item[--]
The variations of the signal and background components in the Monte Carlo 
simulation were allowed ($\Delta Br$ = 0.06).
\item[--]
The upper edge of the decay vertex position was varied along the setup axis
between the chamber PC1 and the calorimeter SP2 ($\Delta Br$ = 0.03).
\item[--]
The electromagnetic showers in the calorimeter SP2 were not used in the photon 
reconstruction ($\Delta Br$ = 0.01).
\item[--]
The particle identification of the secondary track was not used 
($\Delta Br$ = 0.02).
\item[--]
The energy threshold of the selected photons was varied from the value of
0.8 GeV to 2 GeV ($\Delta Br$ = 0.02).
\item[--]
The variations of the angular and mass deviation cuts were applied 
($\Delta Br$ = 0.03).
\item[--]
The event selection was done with different restrictions applied to the relative 
transverse positions of the SP1 photons with respect to the secondary track 
extrapolation ($\Delta Br$ = 0.04).
\item[--]
The allowed intervals for the ratios
$p_{fit}(\pi^0)/p_{meas}(\pi^0)$ and $p_{fit}(\gamma)/p_{meas}(\gamma)$ 
were varied ($\Delta Br$ = 0.05).
\end{itemize}

\section{Summary and conclusion}

\begin{figure}
\centering\mbox{\epsfig{file=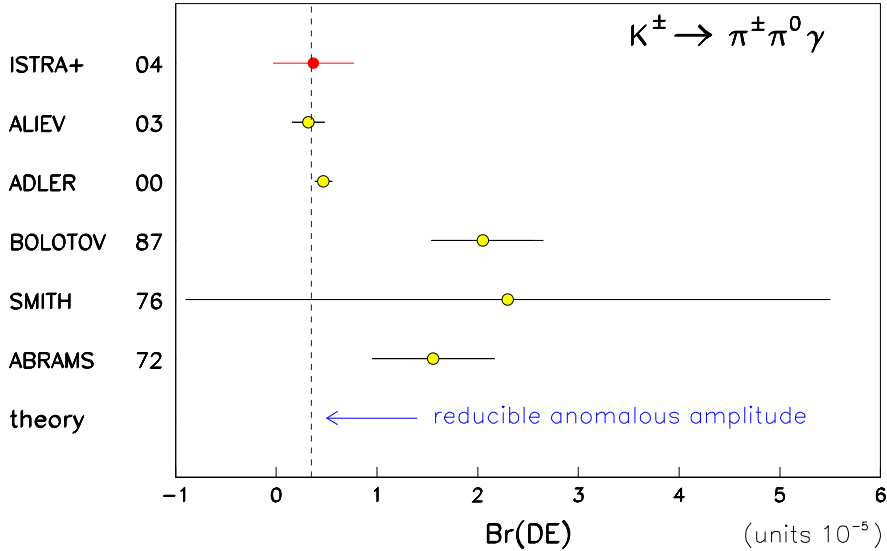,width=0.75\textwidth}}
\caption{\small
The branching ratio for the direct photon emission in the
$K^{\pm} \to \pi^{\pm} \pi^0 \gamma$ decays in the region of
55~MeV\,$< T^*_c <$~90~MeV in comparison with the theoretical prediction under 
the assumption that the direct emission is entirely due to the reducible 
anomalous amplitude \cite{cha2}.}
\label{fig8}
\end{figure}

The branching ratio for the direct photon emission in the
$K^- \to \pi^- \pi^0 \gamma$ decay in the region of 55~MeV\,$< T^*_c <$~90~MeV 
has been measured using the ISTRA+ spectrometer. The results of our measurement 
and the previous experiments [9\,--13] on the 
$K^{\pm} \to \pi^{\pm} \pi^0 \gamma$ decays are presented in Fig.\,\ref{fig8}.
Our value of the branching ratio $Br(DE)$ is consistent with the average value 
of (0.44$\,\pm\,$0.08)$\times$10$^{-5}$, obtained from the results of the
stopped-kaon experiments \cite{Adler00,Aliev03}, but it differs by 2.5 standard
deviations from the average value of (1.8$\,\pm\,$0.4)$\times$10$^{-5}$, 
obtained from the results of the in-flight-kaon experiments [9\,--11].

The theoretical prediction for $Br(DE)$, under the assumption that the direct 
photon emission is entirely due to the reducible anomalous amplitude
\cite{cha2}, is also shown in Fig.\,\ref{fig8} for the same $T^*_c$ region.
Our result supports the hypothesis that the dominant contribution to the direct 
photon emission is due to the pure magnetic transition given by the reducible
anomalous amplitude and the other magnetic and electric amplitudes are small 
or cancelled.

\smallskip
{\it The work is supported by the RFBR grant No. 03-02-16330.}

\end{document}